\documentclass[reprint,
 amsmath,amssymb,
 aps,
floatfix,
]{revtex4-2}

\usepackage{graphicx}% Include figure files
\usepackage{dcolumn}% Align table columns on decimal point
\usepackage{bm}% bold math
\usepackage{float}
\usepackage{mathtools}
\usepackage{makecell,tabularx}
\usepackage{array}
\usepackage{tabularray}
\usepackage{hyperref}
\usepackage{placeins}
\usepackage{titlesec}
\usepackage{lineno}
\usepackage[dvipsnames]{xcolor}

\renewcommand{\selectlanguage}[1]{}
\DeclareUnicodeCharacter{03C8}{$\psi$}

\graphicspath{{Plots/}}

\begin{document}

\preprint{APS/123-QED}

\title{Tagging Efficiency Study of Incoherent Diffractive Vector Meson Production at the Second Interaction Region at the Electron-Ion Collider}

% BNL 2nd EIC detector LDRD PIs
\author{Elke-Caroline Aschenauer}
 %\email{elke@bnl.gov}
\affiliation{
 Department of Physics, Brookhaven National Laboratory, Upton, NY 11973, U.S.A.
}

\author{Alexander Bazilevsky}
 %\email{shura@bnl.gov}
\affiliation{
 Department of Physics, Brookhaven National Laboratory, Upton, NY 11973, U.S.A.
}

\author{Alexander Jentsch}
 %\email{ajentsch@bnl.gov}
\affiliation{
 Department of Physics, Brookhaven National Laboratory, Upton, NY 11973, U.S.A.
}

\author{Jihee Kim}
 \email{jkim11@bnl.gov}
\affiliation{
 Department of Physics, Brookhaven National Laboratory, Upton, NY 11973, U.S.A.
}

\author{Alexander Kiselev}
 %\email{ayk@bnl.gov}
\affiliation{
 Department of Physics, Brookhaven National Laboratory, Upton, NY 11973, U.S.A.
}

\author{Brian Page}
 %\email{bpage@bnl.gov}
\affiliation{
 Department of Physics, Brookhaven National Laboratory, Upton, NY 11973, U.S.A.
}

\author{Zhoudunming Tu}
 %\email{zhoudunming@bnl.gov}
\affiliation{
 Department of Physics, Brookhaven National Laboratory, Upton, NY 11973, U.S.A.
}

\author{Thomas Ullrich}
 %\email{ullrich@bnl.gov}
\affiliation{
 Department of Physics, Brookhaven National Laboratory, Upton, NY 11973, U.S.A.
}

\author{Cheuk-Ping Wong}
 %\email{cwong1@bnl.gov}
\affiliation{
 Department of Physics, Brookhaven National Laboratory, Upton, NY 11973, U.S.A.
}

\date{\today}

\begin{abstract}
%\linenumbers
The Electron-Ion Collider (EIC) is an upcoming accelerator facility aimed at exploring the properties of quarks and gluons in nucleons and nuclei, shedding light on their structure and dynamics. The inaugural experimental apparatus, ePIC (electron-Proton and Ion Collider), is designed as a general purpose detector to address the NAS/NSAC physics program at the EIC. The wider EIC community is strongly supporting a second interaction region and associated second detector to enhance the full science program. In this study, we evaluate how the second interaction region and detector can be complementary to ePIC. The pre-conceptual layout of an interaction region for the second detector offers a secondary focus that provides better forward detector acceptance at scattering angles near $\theta \sim 0$~mrad, which can specifically enhance the exclusive, tagging, and diffractive physics program. This article presents an analysis of a tagging program using the second interaction region layout with incoherent diffractive vector meson production. The potential for the second interaction region to provide improved vetoing capabilities for incoherent events to elucidate the coherent diffractive cross-section is evaluated. The capability to access the coherent diffractive cross-section is of prime importance for studying the spatial imaging of nucleons and nuclei.
\end{abstract}

\maketitle
%\linenumbers

\section{\label{sec:intro}Introduction}
% EIC Big Picture
The Electron-Ion Collider (EIC)~\cite{web_eic_bnl} will be a novel  facility leveraging Deep Inelastic Scattering (DIS) to study the inner structure of nucleons and nuclei. It is designed to ultimately explore the properties of partons in nucleons and nuclei. In particular, it will provide insight on the quark and gluon structure and dynamics, resulting in a greater understanding of the origin of nucleon spin, mass, and confinement mechanisms which are longstanding puzzles in nuclear physics. 

% EIC and Detector 1
The EIC community outlined the physics program of the EIC in its White Paper~\cite{Accardi:1498519}. In order to accomplish the proposed physics program, the EIC will deliver highly polarized electron beams ($\sim 70~\%$) to collide them with ion beams from polarized protons (and some light nuclei), all the way up to unpolarized heavy nuclei like uranium. Moreover, it requires a wide center-of-mass energy range from 30 -- 140~GeV with a high luminosity of $10^{33} - 10^{34} ~\textrm{cm}^{-2}\textrm{s}^{-1}$ to map out the internal structure of nucleons and nuclei in the $x$ and $Q^{2}$ phase space. Along with the accelerator aspects of the facility, the EIC requires a central detector with large rapidity coverage to detect a large fraction of all final-state particles produced in the collisions. 

In order to perform imaging of the partonic structure in nucleons and nuclei, the EIC requires specialized detectors along the beam line in the forward (hadron-going direction) and backward (electron-going direction) regions. These detectors need to be optimally integrated into the lattice of  the interaction region to balance operational feasibility with physics capability. Having large forward and backward acceptance combined with high luminosity enables to study the full three-dimensional momentum and spatial structure of nucleons and nuclei and its evolution in $x$ and $Q^{2}$. The demanding detector requirements and potential technologies for an EIC detector were published in a comprehensive Yellow Report~\cite{ABDULKHALEK2022122447}. The general-purpose detector resulting from this effort, ePIC (electron-Proton and Ion Collider) will be built at the Interaction Point 6 (IP-6), and is designed to study the complete EIC physics program. 

% EIC Detector 2
The EIC is capable of supporting two Interaction Regions (IR-6 and IR-8), but the current EIC project scope includes only IP-6 with the (ePIC~\cite{epic}) detector at the start of EIC operations. Having two general-purpose collider detectors to support the full EIC science program enables cross-checks for potential key results and enables discoveries through complementarity. Additionally, it enables the combining of independent datasets from the two experiments, a demonstrated approach for improving experimental systematic uncertainties. This is crucial for a high-luminosity machine, which will be primarily dominated by systematic uncertainties, rather than statistical fluctuations, for many of the proposed measurements. As a prime example from the HERA facility at DESY in Germany, the most-relevant predecessor to the EIC ~\cite{h12015combinationmeasurementsinclusivedeep, 2010h1zeus}, the complementary technologies of the H1 and ZEUS detectors showed that combining the measured proton cross section data yielded results with uncertainties significantly smaller than the $\sqrt{2}$ reduction assumed by simple combination of datasets. The improvement in precision was driven by not only simply doubling the amount of data collected, but also reducing systematic uncertainties associated with a single detector configuration on the same measurement. Furthermore, the second detector may employ different detector technologies to measure similar final states, but perhaps with a different emphasis, referred to as ``complementarity" in the design and technology choices. It can also focus on different primary physics programs to optimize the overall sensitivity to the full EIC physics scope, and enable topics beyond the initial focus. Operating two interaction regions at the EIC will require sharing luminosity between both detectors at the same center-of-mass energy, but having different features of the interaction regions design integrated. This prohibits an improvement in statistical uncertainties from running two experiments, but still allows for improved systematic uncertainties, as shown by H1 and ZEUS. It should also be noted that the ultimate EIC performance aims for a roughly $10^{2} - 10^{3}$ (depending on beam energy) increase in luminosity compared to HERA, making statistical uncertainties for most measurements the sub-dominant concern.  The second interaction region will have different blind spots in fiducial acceptance relative to the first one caused by the different crossing angle, allowing for better total acceptance combining measurements from the two experiments than a single experiment could achieve. Additionally, the current conceptual design of the second IR provides improved forward detector acceptance at low transverse momentum (low-$p_{T}$) for forward scattered particles, specifically enhancing the exclusive, tagging, and diffractive physics program. Given these arguments, the EIC community~\cite{eicug} is strongly supporting a second detector at the EIC which complements the capabilities of the first detector.

% Figure of EICYR coherent and incoherent processes
\begin{figure}[t]
    \centering
    \includegraphics[width=1\linewidth, height=8cm]{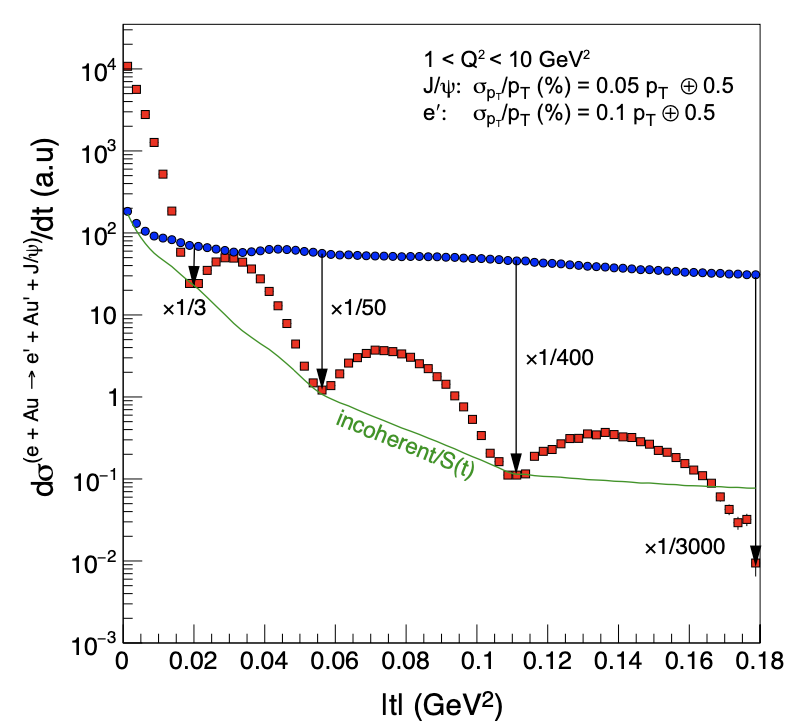}
    \caption{Coherent (red) and incoherent (blue) cross section for diffractive $J/\Psi$ vector meson production in e + Au collisions with the assumed ``nominal" resolution in the range  $1 < Q^{2} < 10$~GeV$^2$. Taken from~\cite{ABDULKHALEK2022122447}}
    \label{fig:eicyr_coherent_incoherent_cross_section}
\end{figure}

% Exclusive Diffractive Vector Meson Measurement - Coherent/Incoherent Processes
In the exclusive, tagging, and diffractive physics program at the EIC, one of the golden measurements is the study of coherent diffractive vector meson production~\cite{Accardi:1498519}, which allows us to study the spatial distribution of partons inside of the nucleus. Unlike Deeply Virtual Compton Scattering (DVCS), vector meson production of $J/\Psi$ mesons provides a tool to specifically image the gluon spatial distribution~\cite{Aschenauer_2019, PhysRevD.106.074019, PhysRevC.87.024913}. With different nuclei, this process can be a test bed for gluon saturation~\cite{MUNIER2001427, PhysRevC.87.024913, PhysRevC.81.025203} at the EIC by measuring different gluon density distributions. Since the vector meson is generated and mostly concentrated in the mid-rapidity region of the central detector, this process is straightforward to measure experimentally. The recoiling target nucleus, or the nuclear breakup products, need to be measured with specialized detectors in the outgoing hadron beam line (``far-forward" detectors) located upto $\sim30$ meters from the interaction point. It is critical to identify the coherent subprocess because diffractive vector meson production contains the sum of coherent (nucleus stays intact) and incoherent (nucleus breaks up) processes, where the incoherent diffractive vector meson production becomes a dominant background to identify the coherent final state. As an example shown in Figure~\ref{fig:eicyr_coherent_incoherent_cross_section}, above $|t|\approx0.015$~GeV$^2$, the incoherent event contribution is dominating and overwhelming the diffractive structure observed from coherent events. Tagging of a coherent nucleus, as opposed to a proton, provides an extreme experimental challenge in all, but the nuclei at very small A. The key approach to extract the coherent diffractive features up the third minima position, is to tag the incoherent processes and veto them with extremely high efficiency ($>99~\%$). Therefore, in order to distinguish coherent from incoherent diffractive production, the experiment requires the ability to tag incoherent diffractive events, where the nucleus breaks up into fragments, which are scattered into the far-forward direction close to the hadron beam line. The specialized detectors in the forward interaction region are positioned to detect nuclear fragments, such as charged hadrons and neutral particles, traveling along with the hadron beam. 

% Figure of Interaction Regions
\begin{figure*}[!hbt]
    \centering
    \includegraphics[width=1\linewidth, height=3cm]{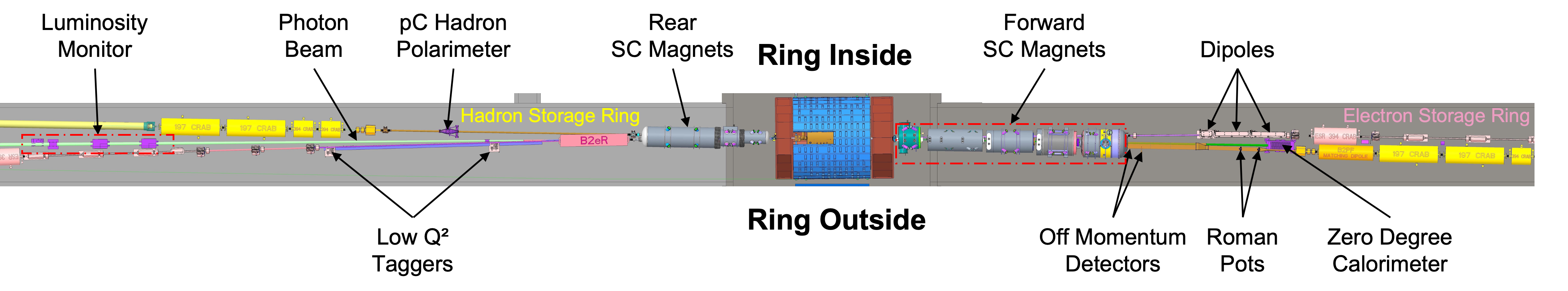}
    \caption{Full layout of the first interaction region (IR-6) with a 25~mrad crossing angle and associated beam line instrumentation.}
    \label{fig:ir6}
\end{figure*}

\begin{figure}[!hbt]
    \centering
    \includegraphics[width=1\linewidth, height=6cm]{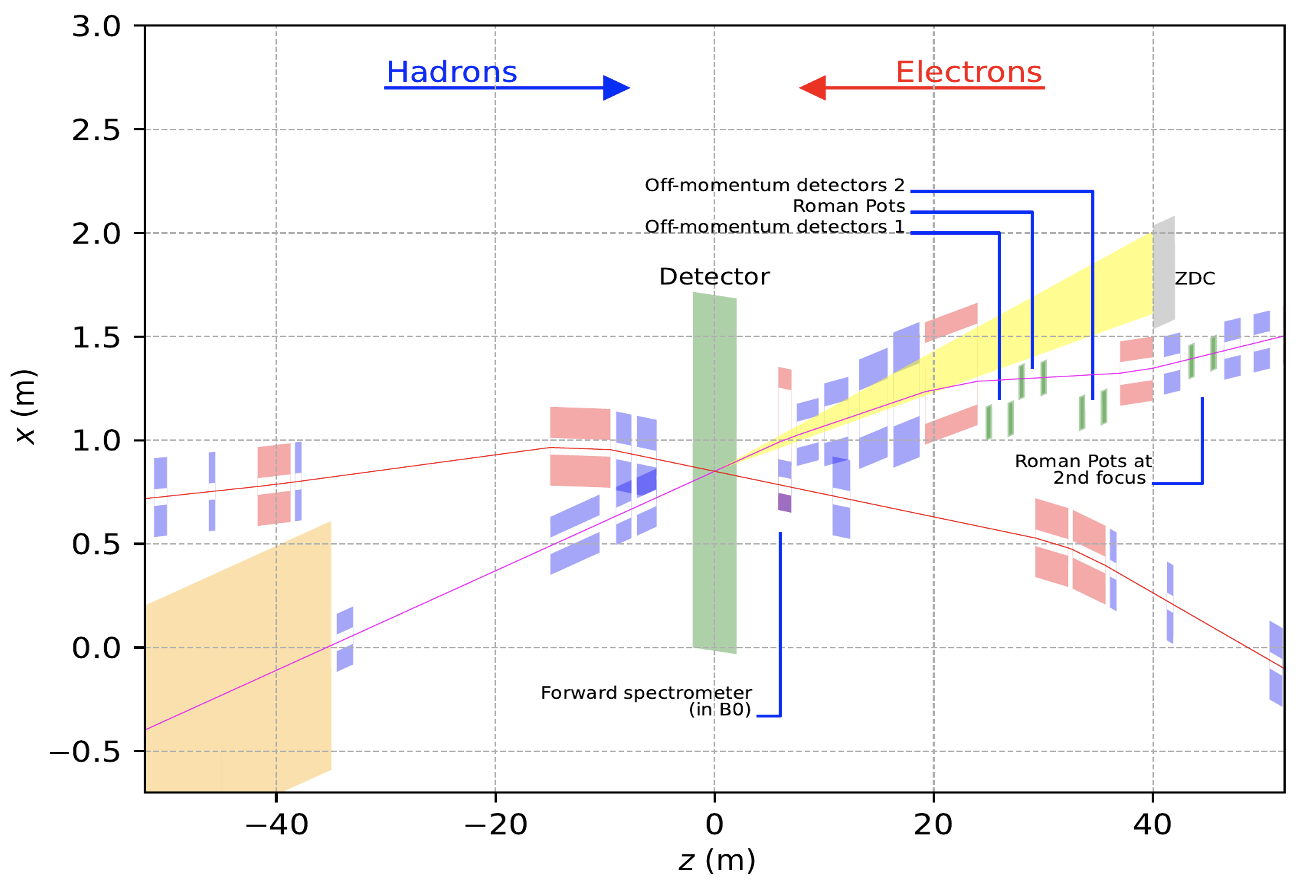}
    \caption{Schematic layout of the second interaction region (IR-8) with a 35~mrad crossing angle with lattice design and potential detector location, and  the secondary beam focus position.}
    \label{fig:ir8}
\end{figure}

% Table of Interaction Comparisons
\begin{table*}[]
\begin{center}
\begin{tabular}{|c|c|c|}
  \hline\hline
    Features & IR-6 & IR-8\\
  \hline\hline
    Accelerator & \multicolumn{2}{c|}{same highlights and challenges}\\
  \hline
    Luminosity & \multicolumn{2}{c|}{shared between both IRs} \\
  \hline
    Center-of-mass energy & \multicolumn{2}{c|}{30 - 140 GeV} \\
  \hline
    Crossing angle & 25 mrad & 35 mrad\\
  \hline
    Transverse momentum & 0.2 $< p_{T} <$ 1.3 GeV & $p_{T} \approx$ 0~GeV (secondary focus)$< p_{T} <$ 1.3 GeV \\
  \hline
    Neutron acceptance & 0.0 $< \theta <$ 4.5 & 0.0 $< \theta <$ 5.5 \\  
  \hline\hline
\end{tabular}
\end{center}
\caption{Comparisons between IR-6 design and pre-conceptual IR-8 design. Both IRs can in-principle have the same coverage at high-$p_{T}$, but the ultimate performance for IR-8 depends on the final apertures of the magnets in the far-forward direction.}
\label{table:ir_comparison}
\end{table*} 

% Brief Goal and Method of This Study
In this article, we use a pre-conceptual design of the EIC second interaction region (IR-8) and explore its capability to identify incoherent eA scattering event with high efficiency. In order to explore new physics opportunities we are taking full advantage of the IR-8 design and choosing detector concepts, which leverage the primary design feature of interest here, namely, a secondary focus. Simulations include the required far-forward detectors adapted to the proposed IR-8 hadron beam line geometry and its present magnetic field configuration. Using an established incoherent event generator (BeALGE~\cite{PhysRevD.106.012007}), the capabilities of separating coherent from incoherent events and separation power added by tagging far-forward nuclear fragments at the secondary focus are evaluated. This program is complementary to the diffractive and tagging physics program at ePIC and will enhance the overall physics capabilities of the EIC.

% Cross References to Each Sections
This paper is structured as follows: Section~\ref{IR8} provides a general description of the proposed beam line for the second EIC detector at IP-8; Section~\ref{FFdetector} describes the implementation of far-forward detectors in the simulation and their expected detector acceptances; Section~\ref{BeAGLE} introduces the event generator being used in this study to determine the tagging efficiency of exclusive incoherent diffractive events using far-forward detectors; Section~\ref{Result} presents the result of the separation power for coherent from incoherent diffractive events by tagging far-forward nuclear fragments; and finally, we summarize in Section~\ref{summary}.

\section{Proposed IR-8 Layout} \label{IR8}
% Interaction Region
The second EIC interaction region~\cite{gamage:ipac2022-mopotk046} shown in Figure~\ref{fig:ir8} shows a complementary design to improve detection capabilities, compared to IR-6 in Figure~\ref{fig:ir6}, for low-$p_{T}$ ($< 200$ MeV) scattered protons and ions at very small angles ($\theta \sim 0$ mrad) as well as to optimize and possible extent the physics reach of the IR-6. Table~\ref{table:ir_comparison} shows a comparison of features between the first and second EIC interaction regions. Similar to the first interaction region, the second interaction region is designed to fit in the existing IP-8 experimental hall and indicates all the beam line detector components in the forward and backward regions. Due to the geometric constraints of the existing experimental hall and tunnel, the second interaction region's optimal crossing angle is proposed to be 35~mrad, which is larger than that of the first interaction region. Due to the larger crossing angle, IR-8 has different blind spots in pseudo-rapidity, and it will be more difficult to get acceptance at high pseudo-rapidity ($\eta \sim 3.5$) in the central detector. However, the proposed hadron beam line for the second EIC interaction region introduces an optical configuration which includes a secondary focus about 45~m downstream of the IP-8 by placing additional dipole and quadruple magnets in the lattice. The basic idea of a secondary focus is to have a very narrow beam profile in the transverse plane, similar to the focus at the interaction point, making it possible to detect particles with very small changes in magnetic rigidity and at very small scattering angle near $\sim 0$~mrad. The second EIC detector can benefit from this machine design as it provides complementarity to the first EIC detector for the exclusive, tagging, and diffractive physics program. 

% Secondary Beam Optics Focus
The placement of silicon detectors around a secondary focus enables the detectors to move closer to the core of the beam, substantially improving the forward detector acceptance for scattered protons or ions from diffractive events and of nuclear fragments, which are otherwise undetectable, because they are too close to, or even within, the beam envelope. In particular, it enables a higher probability to detect low transverse momenta ($p_{T} <$ 200~MeV) particles and to measure particles with a small longitudinal momentum loss (high $x_{L} \sim \frac{p_{\textrm{proton}}}{p_{\textrm{beam}}}$, synonymous with very small magnetic rigidity loss), where these particles are normally very close to the nominal beam orbit and beam total momentum and therefore more difficult to detect. Most importantly, the improved detection of nuclear fragments improves the separation of coherent from incoherent diffractive events, necessary for the 3D imaging program of nuclei, and potentially enables momentum reconstruction of those nuclear fragments. 

% Table of Far-Forward Detector Angular Detectors
\begin{table*}[]
\begin{center}
\begin{tabular}{|c|c|c|}
  \hline\hline
    Detectors & Angular acceptance [mrad] & Notes\\
  \hline\hline
    B0 spectrometer & 5.0 $< \theta <$ 20.0 & tracker and calorimeter\\
  \hline
    Off-momentum detector & 0.0 $< \theta <$ 5.0 & 40-60\% rigidity\\
  \hline
    Zero-degree calorimeter & $\theta <$ 5.5 & neutral particles full acceptance upto 5~mrad\\
  \hline
    Roman Pot detector & 0.0 $< \theta <$ 5.0 & 10$\sigma$ safe-distance cut\\
  \hline\hline
\end{tabular}
\end{center}
\caption{Summary of angular acceptance for the four far-forward detectors for the pre-conceptual IR-8 design, assuming a very similar concept to the IR-6/ePIC design.}
\label{table:ff_acceptance}
\end{table*}

\section{Far-Forward Detectors} \label{FFdetector}
\begin{figure}[t]
    \centering
    \includegraphics[width=1\linewidth, height=4cm]{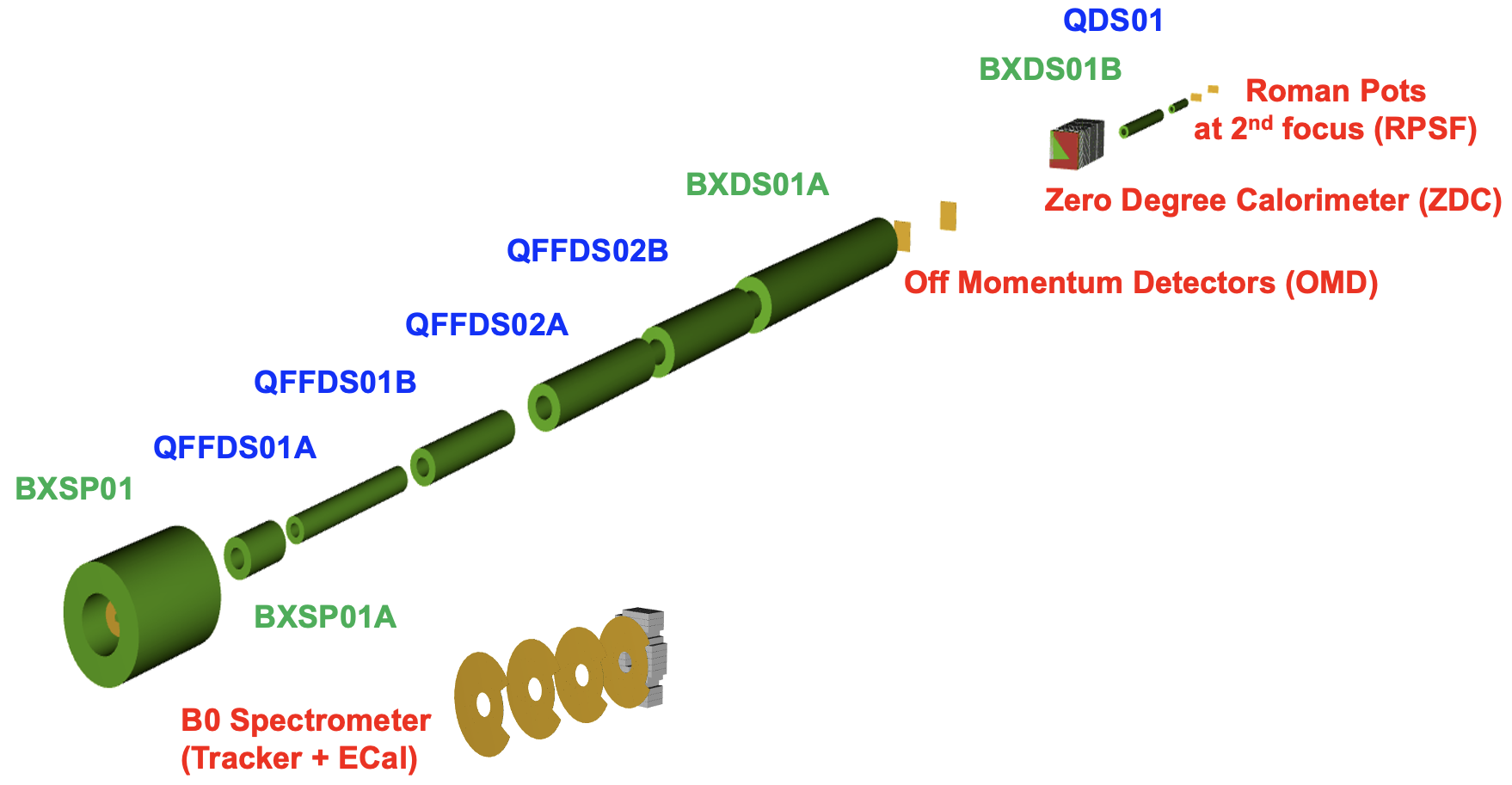}
    \caption{The layout of the second IR far-forward region in the DD4hep simulation. It includes four dipole (green), five quadrupole (blue) magnets in the ion beamline, and four far-forward detector subsystems. The image was generated using the DD4hep simulation.}
    \label{fig:ir8_ff}
\end{figure}

% Detector Layout
For diffractive physics measurements, the EIC detector is required to cover a broad acceptance for scattered charged and neutral particles not only in the central detector, but also in the far-forward region. It is important to integrate specialized detectors into the interaction region lattice to fulfill the needs of diffractive physics measurements. The IR integrates multiple detectors along the beam line, comprised of trackers and calorimetry, to measure their momenta and energy. The main purpose of the far-forward detectors is to tag scattered, intact neutrons, protons or ions from diffractive events, and to tag nuclear fragments from various breakup events. To explore new physics opportunities by taking full advantage of this secondary focus in IR-8, we included a pre-conceptual implementation~\cite{D2EIC_github} of the required far-forward detectors with the proposed beam line geometry and its field configuration, which is shown in Figure~\ref{fig:ir8_ff}. The current detector layout is fully integrated in a DD4hep (Detector Description for High Energy Physics)~\cite{frank_markus_2018_1464634} based simulation of the described IR-8 design. The Roman pot detectors are placed at the secondary focus location with the general far-forward detector configuration being very similar to the one in IR-6. The angular acceptances of these far-forward detectors are described in Table~\ref{table:ff_acceptance}.

% Figure of Detector Layout in DD4hep Simulation

The far-forward specialized detector systems consists of the B0 spectrometer, a tracker and an electromagnetic calorimeter (EMCAL), the Off-Momentum Detector (OMD), a Zero-Degree Calorimeter (ZDC), and Roman Pot (RP). The general features of each detector subsystem are as follows.

\begin{description}
% B0 Spectrometer
\item[B0 spectrometer] A subsystem designed to detect scattered protons and photons with relatively larger angles between 5~mrad and 20~mrad compared to the particles in the far-forward region. It consists of four silicon tracking layers and followed by 10~cm lead-tungsten homogeneous crystal calorimeter, which uses 20~cm length of crystal calorimeter in ePIC. The entire detector system is sitting in the bore of the 1.3~T BXSP01 dipole magnet, where the B0 name is an adoption of the name used in the IP-6 configuration, where the equivalent magnet is labeled ``B0" (the zero-th bending magnet after the IP). Using four tracking layers in a magnetic field enables full momentum reconstruction of charged particles using conventional tracking methods.

% Off-Momentum Detector
\item[Off-Momentum Detector] This subsystem is designed to detect scattered protons and charged particles from nuclear breakups with below 60~\% of the beam particles' rigidity. They consist of two silicon tracking layers, placed 2~meters apart and covering a scattering angle acceptance of 5~mrad in the outgoing hadron beam line.

% Zero-Degree Calorimeter
\item[Zero-Degree Calorimeter] This system is designed to detect high energy neutrons and photons, and low energy photons ($> 100~\textrm{MeV}$) relevant for incoherent/coherent event separation. Neutral trajectories are not bent by magnets and therefore travel in a straight line from the IP along the crossing angle value for the IR. The ZDC consists of both an electromagnetic calorimeter and a hadronic calorimeter (HCAL). The EMCAL consists of a crystal (PbWO$_4$) and a silicon pixel layer while the HCAL consists of scintillator plates interleaved with silicon and lead plates (Pb/Si and Pb/Sci) corresponding to seven interaction lengths. Note that the ZDC technologies used in this simulation is the old design used for IR-6, which now uses a combination of a PbWO$_4$ crystal EMCAL and a SiPM-on-tile imaging HCAL inspired by the design of CALICE concept~\cite{CAdloff_2012}.

% Roman Pot Detector and 10$\sigma$ Safe Distance Cut
\item[Roman Pot] The Roman pots are designed to detect scattered protons and nuclear fragments with very small changes in their rigidity up to 5~mrad in scattering angle. They have two silicon tracking layers positioned at the secondary focus location so that one can see scattered particles with $p_T$ close to 0. The displacement from the core of the beam is determined by a 10$\sigma$ safe-distance requirement, which is calculated based on beam parameters as described in Equation~\ref{eq:10sigmacut}.
% Equation of transverse beam size
\begin{equation}
    \sigma_{x,y} = \sqrt{\epsilon_{x,y}\beta(z)_{x,y} + (D(z)_{x,y}\frac{\Delta p}{p})^2}
    \label{eq:10sigmacut}
\end{equation}
where $\sigma$ represents the transverse beam size, $\epsilon$ represents the beam emittance, which is a constant value around the ring, $\beta$ represents the beta function, which describes the transverse size of the beam at location $z$ along the nominal beam trajectory, $D$ represents the momentum dispersion at location $z$ along the nominal beam trajectory, and $\frac{\Delta p}{p}$ represents the beam momentum spread. Some parameters describing the beam are from the EIC CDR~\cite{Willeke:2021ymc} and some beam parameters are evaluated from a recent IR-8 simulation~\cite{Randy_2023}. At the secondary focus, $10\sigma_{x}$ and $10\sigma_{y}$ were calculated to be 1.5 millimeters and 1 millimeters, respectively, for the top-energy (110~GeV/n) heavy-ion beams.
\end{description}

\section{Event Generator - BeAGLE} \label{BeAGLE}
% Nuclear Fragments at Generator Level
\begin{figure*}[]
    \centering
    \includegraphics[width=1\linewidth, height=8cm]{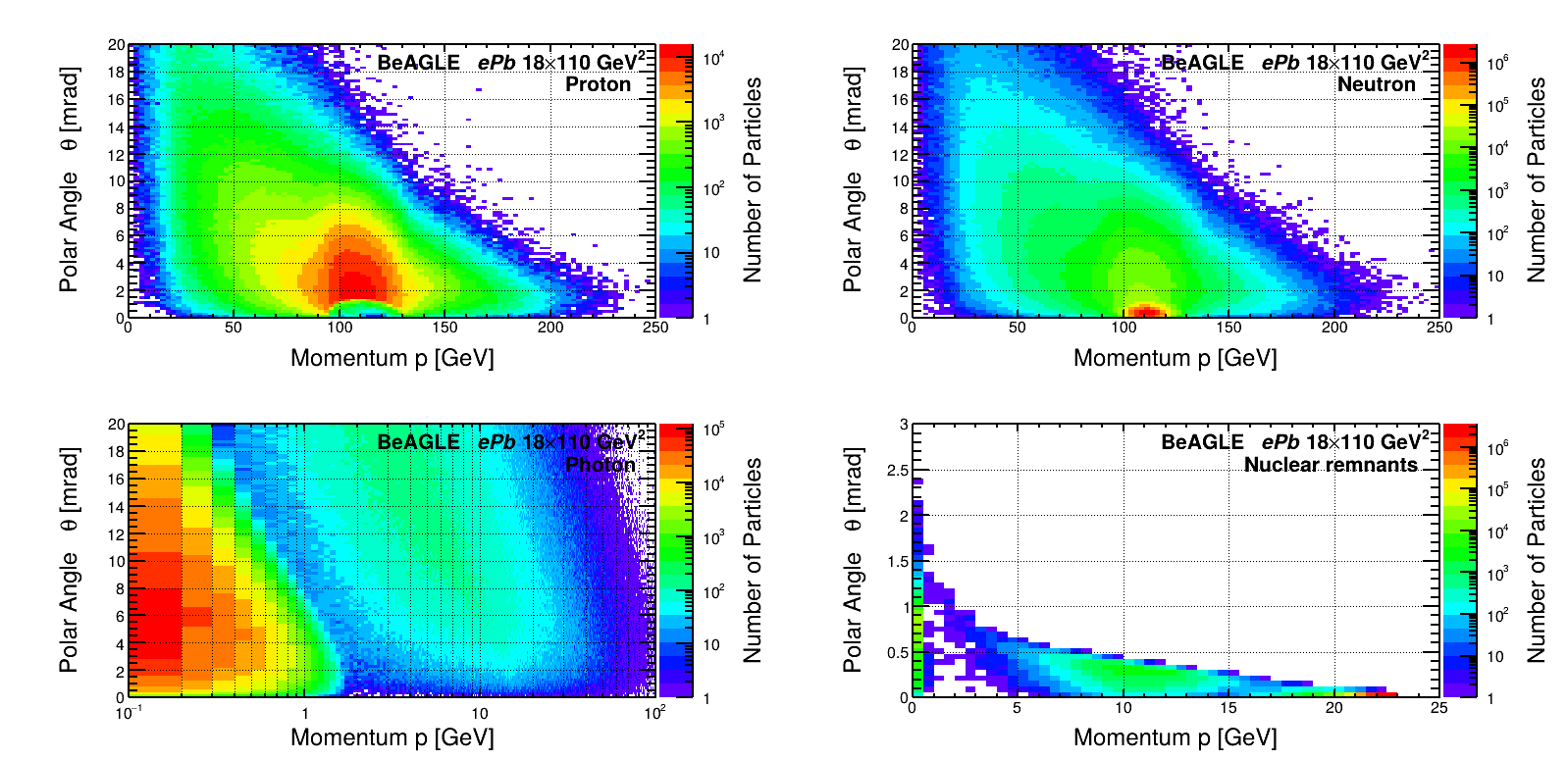}
    \caption{The polar angle distribution as a function of each nuclear fragment momentum for protons, neutrons, photons, and nuclear remnants are at the generator level.}
    \label{fig:ptl_generator_level}
\end{figure*}

% BeAGLE
In order to evaluate the impact on the tagging efficiency based on the pre-conceptual design of IR-8 including the secondary focus and the far-forward detectors, we used the BeAGLE~\cite{PhysRevD.106.012007} event generator to generate incoherent events of vector meson production: $e + Pb \rightarrow e' + J/\Psi + X$. BeAGLE is an established eA incoherent event generator and is extensively used to investigate detector performance at the EIC~\cite{TU2020135877, PhysRevC.104.065205, magdy2024studynuclearstructurelight}.

% Sample for this Study
For this study, we ran the BeAGLE event generator version 1.03.02 to create a sample of $e + Pb \rightarrow e' + J/\Psi + X$ with 18~GeV electrons colliding with 110~GeV per nucleon lead nuclei. The $1\times10^{7}$ events were passed through the EIC afterburner~\cite{afterburner_github} to properly introduce the crossing angle and beam effects, such as angular divergence and momentum spread. This ensures that the simulation is realistic as the forward particle acceptance and momentum resolution are especially sensitive to these effects. Figure~\ref{fig:ptl_generator_level} shows the two-dimensional distributions of scattering angle versus the momenta of protons, neutrons, photons, and nuclear fragments being generated in the process of incoherent vector meson ($J/\Psi$) production in ePb collisions. As seen in Figure~\ref{fig:ptl_generator_level}, the majority of protons are within a scattering angle less than 5~mrad and are expected to be tagged by the OMD, neutrons within a 5~mrad scattering angle are detected by the ZDC, photons are expected to be captured by a combination of the B0 EMCAL and the ZDC, and the heavy nuclear remnants can be detected by the RP, this is a particular capability enabled by the IR-8 design concept. Table~\ref{table:ptl_generator_level} shows any combination of nuclear breakup particles from incoherent diffractive vector meson ($J/\Psi$) production in ePb collisions at the BeAGLE generator level. Based on the generator level, about 95~\% of incoherent diffractive events have some number of neutrons while the rest have fragments, which can be tagged by RP at the secondary focus. 

% Table of combinations of nuclear breakups at final-state at generator level
\begin{table}[]
\begin{center}
\begin{tabular}{|c|c|}
  \hline\hline
    Nuclear breakups at final-state & Number of events\\
  \hline\hline
    Only nuclear fragments & 0.0005 \% \\
  \hline
    + neutrons & 7.8458 \% \\
  \hline
    + protons & 0.0001 \% \\
  \hline
    + photons & 3.4132 \% \\
  \hline
    + neutrons + protons & 3.2034 \% \\
  \hline
    + neutrons + photons & 45.4347 \% \\
  \hline
    + protons + photons & 1.8598 \% \\  
  \hline
    + neutrons + protons + photons & 38.2426 \% \\      
  \hline\hline
\end{tabular}
\end{center}
\caption{Summary of combinations of final-state nuclear breakup particles at the generator level.}
\label{table:ptl_generator_level}
\end{table}

\section{Result} \label{Result}
% Vetoing Power
\begin{figure*}[]
    \centering
    \includegraphics[width=\linewidth]{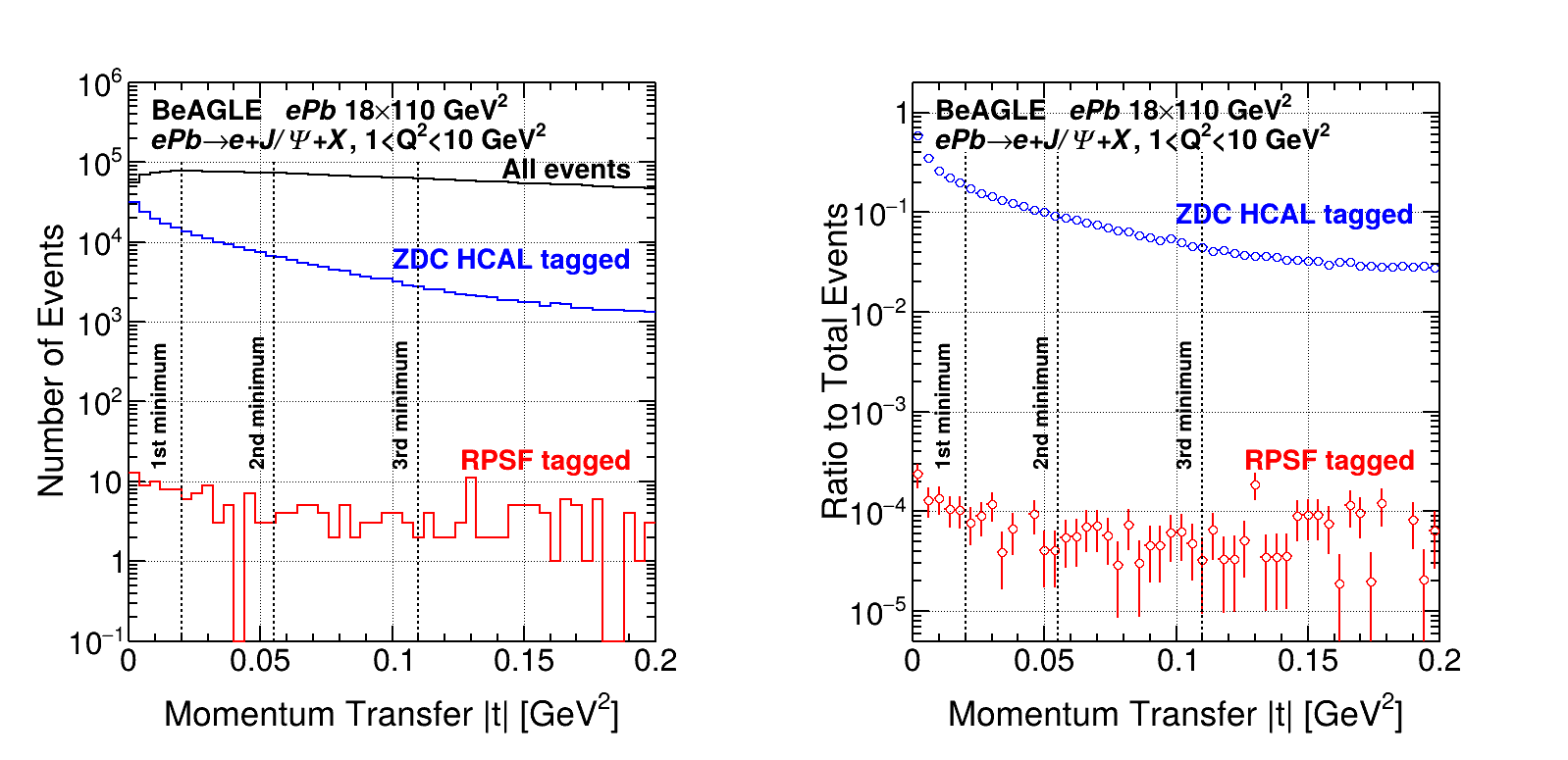}
    \caption{Left: The number of non-vetoed incoherent diffractive events for ePb collisions as a function of $t$. The black line represents all incoherent events, the blue line represents surviving events after ZDC tagging and vetoing, and red line represents surviving events after ZDC and RP tagging and vetoing. Right: The vetoing power distribution as a function $t$. Each line represents each veto inefficiency histogram divided by the total incoherent histogram. The negligible impact on the other vetoing selections are not shown here. Note that the term ``RPSF" refers to the Roman Pot located at the Secondary Focus.}
    \label{fig:vetoing_power}
\end{figure*}

% Intro and Tagging/Vetoing Procedure
In this section, the impact of the secondary focus in the pre-conceptual IR-8 design to suppress the incoherent diffractive contribution is examined, which is the dominant background to coherent diffractive $J/\Psi$ production. While reconstruction of the very forward scattered protons from Deeply Virtual Compton Scattering (DVCS) at the B0 spectrometer and/or Roman pot detectors requires hit information from all tracker layers for full track reconstruction, for studying the vetoing efficiency of incoherent diffractive events, it is enough to tag at least one of the nuclear breakup particles from activity in a single detector layer to identify an incoherent event. Note that this should be reevaluated with realistic detector background simulations in the future. The detailed veto selections are as follows:

% Veto Procedure
\begin{itemize}
    \item Veto 1: Any registered hits in the ZDC HCAL
    \item Veto 2: Any registered hits by one layer of RP located closest to the secondary focus, and hits should be outside 10$\sigma$ safe-distance cut
    \item Veto 3: Any registered hits of two layers out of the OMD
    \item Veto 4: Any registered hits in at least two out of the four layers of the B0 tracker
    \item Veto 5: Energy deposit of all registered hits with greater than 100~MeV in the B0 EMCAL
    \item Veto 6: Energy deposit of all registered hits with greater than 100~MeV in the ZDC EMCAL
\end{itemize}

% Effect of Secondary Focus on Tagging
The left panel of Figure~\ref{fig:vetoing_power} shows the number of non-vetoed incoherent diffractive events of ePb collisions and the right panel of Figure~\ref{fig:vetoing_power} shows a ratio of non-vetoed events to total incoherent events of ePb collisions. The tagging impact of particularly the ZDC and RP at the secondary focus is highlighted as a function of $t$ for the incoherent events. As seen in Figure~\ref{fig:vetoing_power}, a large vetoing impact is observed from the ZDC to the RP at the secondary focus. Unlike for IR-6~\cite{PhysRevD.104.114030}, heavy nuclear fragments can be tagged by the RP much more often. This is due to the fact that the beam size at the secondary focus is nearly as small as at the interaction point, enabling placement of the RP closer to the beam core. This additional tagging capability, enabled by the IR-8 design, has a drastic impact on the vetoing efficiency for incoherent events. The EIC Yellow Report~\cite{ABDULKHALEK2022122447} describes the needed rejection factor of incoherent events as a function of $t$ to make the coherent diffractive ``dips"  visible, as shown in Figure~\ref{fig:eicyr_coherent_incoherent_cross_section}. This is especially challenging at the position of the third diffractive minimum of the coherent diffractive pattern of $J/\Psi$ production, where the incoherent cross section is about 400 times larger than the coherent cross section, meaning a suppression factor of incoherent events better than 400:1 must be achieved to distinguish between coherent and incoherent diffractive processes. In the IR-6 design~\cite{PhysRevD.104.114030}, the achieved rejection of incoherent events is enough to see the first minimum but not the second or third minima. Comparing to the IR-6, the proposed IR-8 design is found to enable suppression of the incoherent diffractive contribution at all three minima and the vetoing power is greater than $10^{3}$ for the current geometry integrated in the simulation without implementing a beam pipe as well as no background and detector noise. Figure~\ref{fig:vetoing_power} demonstrates that tagging at the secondary focus provides a stronger veto at any $t$, with $10^{3}$ times of rejection power. After all veto selections, the surviving events consist of nuclear fragments with high atomic mass number (A = 208 and 207 for Pb) and low particle multiplicity. These final-state nuclear fragments (i.e. A-1 particles) are very close to the beam momentum and remain likely within the beam envelope, making detection impossible, independent of the IR design.

% Beam Pipe Study
In order to evaluate the acceptance impact of the beam pipe on the vetoing efficiency, we implemented a simplified version of the beam pipe (following the present design concept for IR-6) up to the neutron exit window for the ZDC. The position of the neutron exit window was evaluated from a single particle gun simulation with a 5~mrad cone, where neutrons and protons begin to be separable from each other. In addition to the aperture, the material, the thickness, and tilt of the exit window with respect to the neutral particle axis affects the neutron detection of the ZDC.  It is anticipated not to have a major effect on the total vetoing efficiency, however, because the main gain in vetoing power comes from the veto of heavy nuclear fragments using the secondary focus. After the beam pipe and neutron exit window were implemented, the neutron acceptance is reduced from 5~mrad to 4.5~mrad, however, each event has multiple neutrons and nuclear fragments that can be captured at the RP at the secondary focus. Table~\ref{table:veto_eff_beam_pipe} shows a comparison on the number of surviving events with and without implementing a beam pipe in the same sample. The vetoing power is not affected much with respect to the inclusion of the beam pipe. It is important to note that the current IR-8 design is at the very initial stage and more detailed machine studies are needed to understand how to operate the EIC with two IRs, while we get the best and stable operation in general. 

% Table of Surviving Events with/without Beam Pipe
\begin{table}[h!]
\begin{center}
\begin{tabular}{|c|c|c|}
  \hline
  \multicolumn{3}{|c|}{Number of surviving events} \\
  \hline\hline
    Veto Selections & Without beam pipe & With beam pipe\\
  \hline\hline
    All events & 100 \% &  100 \% \\
  \hline
    ZDC HCAL tagged & 5.7134 \% &  5.7971 \% \\
  \hline
    RPSF tagged & 0.0097 \% &  0.0090 \% \\
  \hline
    OMD tagged & 0.0096 \% &  0.0087 \% \\
  \hline
    B0 tracker tagged & 0.0041 \% &  0.0041 \% \\
  \hline
    B0 ECAL tagged & 0.0023 \% &  0.0026 \% \\
  \hline
    ZDC ECAL tagged & 0.0005 \% &  0.0014 \% \\    
  \hline\hline
\end{tabular}
\end{center}
\caption{Summary of the number of events surviving before/after beam pipe implementation at the different veto cuts.}
\label{table:veto_eff_beam_pipe}
\end{table}

\section{Summary and Outlook} \label{summary}
% Summary
In this article a study of the vetoing efficiency for incoherent diffractive vector-meson production using a sample of 18~GeV electrons on 110~GeV per nucleon lead nuclei from the BeAGLE event generator for the pre-conceptual IR-8 design is presented. In the design of the second interaction region, the inclusion of the secondary focus enables substantial improvement of the forward detector acceptance of the scattered protons and light ions at very low $p_{T} \sim 0$~GeV/c, and of nuclear fragments close to beam-rigidity overall. This secondary focus provides an advantage for tagging particles with low $p_{T}$ and/or high $x_{L}$. This allows for tagging of scattered particles with high rigidity which are traveling essentially collinear to the beam particles themselves. Most importantly, the improved detection of nuclear fragments makes it possible to distinguish coherent from incoherent diffractive events over a broad range of momentum transfer. This simulation result shows that the pre-conceptual design of the EIC second interaction region can be highly complementary to the first detector and offers unique capabilities to enhance the overall EIC exclusive, tagging, and diffractive physics program.

% Outlook
The results shown in this paper are based on the proposed pre-conceptual design of IR-8 at the EIC. Note that the far-forward detector acceptances are significantly affected by the machine design and need to be reassessed periodically as the design progresses and matures. Although a more detailed and realistic IR-8 design is needed in the future to further evaluate the benefits of having the secondary focus, the current results show an opportunity to enhance the exclusive, tagging, and diffractive physics program with improved far-forward acceptance for such an IR-8 design. This will impact different exclusive physics processes and provide the ability to cross check measurements from IP-6, and enable complementary fiducial acceptances between the various far-forward detectors between the two IRs. Additionally, along with improving tagging performance, it may be interesting to study possibilities to directly tag coherent diffraction events in a range of medium A ions available at the EIC. Future plans will be to study various ion beams from light to medium A-nuclei to explore the capabilities/limits/acceptances on coherent tagging at the secondary focus, where it can provide complimentary measurements, as well as unique capabilities offered from the second interaction region. This study provides the foundation needed to refine the IR-8 design and enable detailed study of quark and gluon dynamics in nuclei which will be relegated to a future physics analysis with a better-developed conceptual design.

\section*{\label{sec:Acknowledgements} Acknowledgements}
We would like to thank Xiaoxuan Chu, Oleg Eyser, Akio Ogawa, Zhengqiao Zhang and the local BNL group for helpful discussion on this topic. I would like to thank Bamunuvita Randika Prasad Gamage for discussion on the IR-8 lattice design and simulation. Also, I would like to thank Renee Fatemi, Pawel Nadel-Turonski, Anna Stasto, and Wenliang Li for general discussions on the EIC detector 2. The work is supported by Laboratory Directed Research and Development (LDRD) funding, ``A Second EIC Detector: Physics Case and Conceptual Design" from Brookhaven National Laboratory under LDRD-23-050 project.

\bibliography{main}% Produces the bibliography via BibTeX.

\end{document}